# FR-Detect: A Multi-Modal Framework for Early Fake News Detection on Social Media Using Publishers Features

Ali Jarrahi[1]

*Computer Engineering, University of Zanjan, Zanjan, Iran*
*jarrahi@znu.ac.ir*

Leila Safari

*Computer Engineering, University of Zanjan, Zanjan, Iran*
*lsafari@znu.ac.ir*

**Abstract**

In recent years, with the expansion of the Internet and attractive social media infrastructures, people prefer to follow the news through these media. Despite the many advantages of these media in the news field, the lack of any control and verification mechanism has led to the spread of fake news, as one of the most important threats to democracy, economy, journalism and freedom of expression. Designing and using automatic methods to detect fake news on social media has become a significant challenge. In this paper, we examine the publishers' role in detecting fake news on social media. We also suggest a high accurate multi-modal framework, namely FR-Detect, using user-related and content-related features with early detection capability. For this purpose, two new user-related features, namely *Activity Credibility* and *Influence*, have been introduced for publishers. Furthermore, a sentence-level convolutional neural network is provided to combine these features with latent textual content features properly. Experimental results have shown that the publishers' features can improve the performance of content-based models by up to 13% and 29% in accuracy and F1-score, respectively.

## 1- Introduction

Nowadays, the Internet has made up a significant part of human lifestyle. The role of traditional news channels, such as newspapers and television, has been diminished and weakened greatly in the news reception. In particular, the expansion of social media infrastructures, such as Facebook and Twitter, have had a significant role in undermining traditional media. People use social media to connect with friends, relatives and gather information and news from around the world. The reason for this behavior can be traced back to the nature of these media. First, it is much faster and less expensive to get news through these media than traditional media. Second, it is easy to share news with friends and other people for further discussions. As of August 2018, around 68% of Americans received news via social media, compared to 62% in 2016 and 49% in 2012[2].

---

[1] *Corresponding author*
[2] https://www.journalism.org/2018/09/10/news-use-across-social-media-platforms-2018/

However, these benefits of social media are not costless. Mainly, the lack of any control and verification of the news releases has made social media a fertile ground for disseminating false or unverified information [1]. Often an attractive news headline is enough for an article to be shared thousands of times despite its inaccurate or unapproved content.

Fake news is not a new phenomenon. In fact, prior to the advent of the Internet, journalists had also been investigating and verifying their news and sources [2], but at that time the impact of fake news on public opinion was very limited and therefore insignificant. Today, with the expansion of social media, the spread of inaccurate or unverified information among a large number of people, regardless of geographical boundaries, has been facilitated. As a result, public perceptions of events can be profoundly affected by fake news [1]. The 2016 US Presidential Election is one of the prominent examples of the impact of spreading fake news [3].

Fake news is now recognized as one of the most important threats to democracy, journalism and freedom of expression, which can even undermine public confidence in governments [4]. The economy is also not immune to the spread of fake news. Significant fluctuations occur with the propagation of fake news related to the stock market [5]. The importance of fake news in recent years has led to the term "fake news" being chosen as the word of the year by Macquarie and Oxford dictionaries in 2016.

Social and psychological factors play an important role in gaining public trust and further spreading fake news. For example, it has been shown that when humans are overly exposed to deceptive information, they become vulnerable and irrational in recognizing truth and falsehood [6]. Studies in social and communication psychology have also shown that human ability to detect deception is only slightly better than chance, with a mean accuracy of 54% obtained over 1,000 participants in over 100 experiments [7]. This situation is more critical for fake news because of its special features. Therefore, it is essential to provide methods for automatic detection of fake news on social media. So, the aim of this paper is to provide a model that can detect fake news with high accuracy as soon as possible. For this, some important publishers' features have been introduced and used alongside content-related features.

The paper is structured as follows. The related concepts to study fake news on social media are presented in the next section. The previous works have been summarized in section 3. The details of the proposed methods are described in section 4. We have evaluated our approach on a comprehensive fake news detection benchmark dataset. The experimental results are presented in section 5. Finally, the paper concludes with future research directions in section 6.

**2- Fake news on social media**

This section provides concepts and definitions related to fake news on social media to give readers and researchers a better understanding of its features. Although there is no comprehensive definition for fake news [4], but a clear definition can help in distinguishing related concepts and better analyzing and evaluating fake news. The definition of news in the Oxford Dictionary is as follows: new information about something that has happened recently. In the context of social media, the most related concept to fake news is rumor. A rumor can be defined as an unverified claim or information, which is created by users on social media and can potentially spread beyond their private network [8]. This unverified information could be true, partly true or completely false, or even remain unverified [1]. Similar to fake news, spreading false rumors can cause serious damages, even in a short period of time.

Researchers in [4] have distinguished related terms and concepts, like rumor and satire news, based on three characteristics: Authenticity (false or not), Intention (bad or not), and Type of information (news or not). For example, a rumor is an information that all these characteristics are *unknown*, whereas fake news is a false news that is presented with a bad intention to mislead the general public or a particular group. So, fake news can be defined as follows: fake news is *intentionally* and *verifiably* false news published by a news outlet [4, 9]. According to the definitions and the characteristics provided, the relationship between the concepts of news, fake news and rumors can be considered as Figure 1.

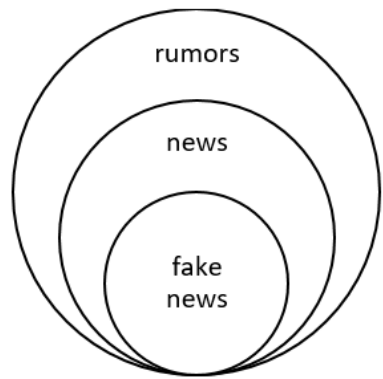

*Figure 1- The relationship between the concepts of news, fake news and rumors.*

In addition to definitions, determining the life cycle of fake news and its related components in social media is essential for the proper study of fake news in this context. Zhou et al. [4] have considered the life cycle of fake news based on three stages: creation, publication and propagation. However, given that fake news is verifiable, we believe that there is a detection stage in the life cycle and eventually all fake news is detected. Therefore, we have modified the life cycle of fake news as shown in Figure 2. Each stage of the life cycle is described below.

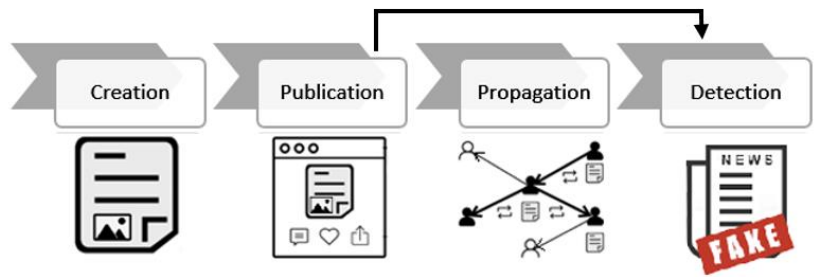

*Figure 2- The life cycle of fake news on social media.*

**Creation:** At this stage, fake news content is created by one or more authors for specific purposes. Creating fake news can be done in the context of social media or outside. The main parts of the news include the headline and the body. There may be other optional sections such as images, authors and news sources.

**Publication:** After creating fake news, it is necessary to inject the created news in social media by one or more publishers. Here, the publisher is actually a user of that social media. Each user on social media has a specific identity that can be defined through features such as friends, followers, history of activities and etc. The published news on social media is primarily received by the followers of each publisher. This stage is called the publication phase.

**Propagation:** After the publication stage, each news article enters a phase that depends entirely on the behavior of the recipients. After receiving the news, each recipient may share, comment or like the news or leave it without any action. In general, the news recipients can be divided into three categories:

- *Malicious User:* A user who intentionally endorses and shares fake news for specific purposes while being aware that the news is fake.
- *Conscious User:* A user who carefully tries to avoid sharing fake or suspicious news as much as possible.
- *Naïve User:* A user who unintentionally shares fake news due to the deception of malicious users and social effects. Naïve users participate in the fake news propagation process because of their prior knowledge (as expressed by *confirmation bias*[3] *[10]*) or the peer-pressure (as explained by *bandwagon effect*[4] *[11]*).

After some news recipients share the fake news, their followers also receive fake news and this process continues. This stage is called the propagation phase.

**Detection:** As stated in the fake news definition, the authenticity of the news can be verified using existing evidence and therefore its falsity can be detected. Of course, it will take a while to determine if the news is fake. The longer this period lasts, the more people on social media will be affected by the fake news. Therefore, it is very important that the detection be made as soon as possible, which is known as early detection in the fake news field. Detection of fake news on social media can be done after the propagation stage or ideally before it. After the news is detected as fake, the propagation phase ends.

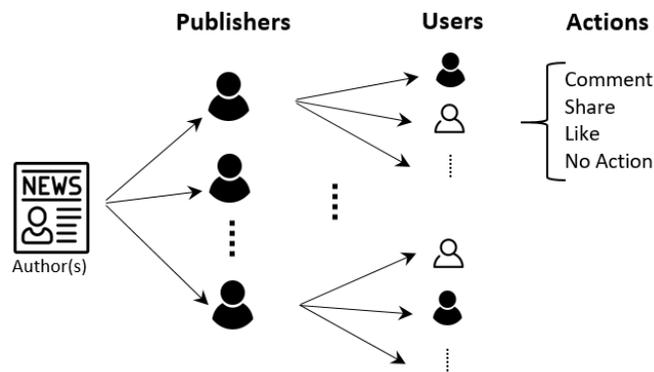

*Figure 3- The process of spreading fake news on social media.*

The process of spreading fake news on social media is summarized in Figure 3 and an example fake news on Facebook is shown in Figure 4. Given the process and its components, there are useful features that can help fake news detection. As summarized in Figure 5, these features can be divided into four general categories, which are described below.

---

[3] Individuals tend to trust information that confirms their preexisting beliefs or hypotheses.
[4] Individuals do something primarily because others are doing it.

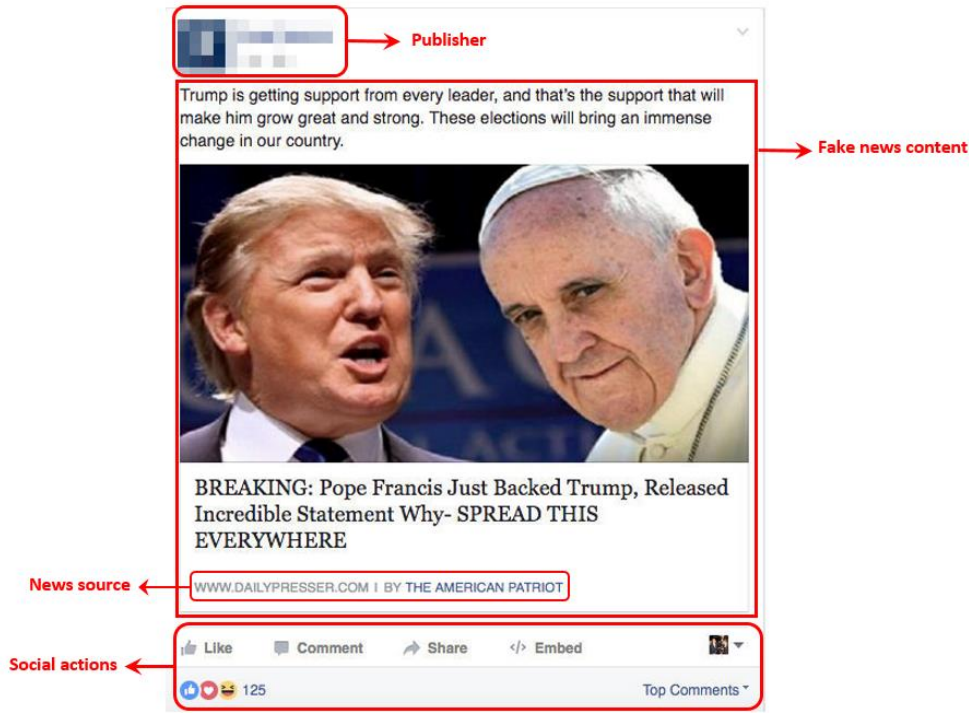

*Figure 4- An example of fake news on Facebook.*

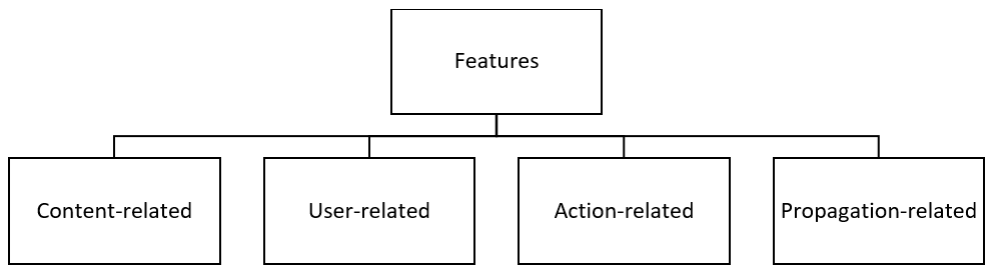

*Figure 5- Types of features available in the fake news life cycle on social media.*

**Content-related features:** Some features are directly related to the news content. Structurally, a news story includes headline, body, image(s), source and author(s). Each of these parts or the relationships between them may contain useful features that can be extracted and utilized.

Writing style features can be used to determine the author's intent (bad or not) [4]. These features can be extracted based on existing theories, such as the complexity of the text (e.g., the average number of words in sentences) and the features that measure the sentiment of the text (e.g., the amount of positive and negative words), Or features extracted from the structure of the text, e.g., *bigram [12]*, *POS* (Part of Speech) [13], *LIWC* (Linguistic Inquiry and Word Count) [12, 14] and *RR* (Rhetorical Relations) [15].

Regarding the writing style features, it is important to note that fake news is generally about important events with financial or political benefits. Therefore, its authors are so motivated to write the news in such a way that it is not detectable by current fake news detection methods. Therefore, developing a real-time representation and learning of writing style features is essential. Deep learning methods can help extracting the latent features in the news content. Therefore, recent writing style based fake news detection methods are mainly rely on deep learning techniques [16, 17].

Other news content features include image-related features such as image forgery and how the image relates to the news body. Another feature is the headline credibility and its relevance to the news body, which is similar to the *clickbait* recognition problem. Authors' credibility as well as news sources can also be helpful in detecting fake news. Analyzing the content of fake news is not sufficient to create an effective and reliable identification system. So, other important aspects, such as the social context information of news, should also be considered [18].

**User-related features:** A social media user, regardless of name or account, is an identity associated with a human or robot that interacts with other users and components in social media. Users have very important features that can be used in fake news detection. Some of these features are listed below:

- *Validity:* This feature indicates whether the user matches the original identity associated with him/her in the real world or not.
- *Lifetime:* This feature indicates the time elapsed since the creation of the user on social media.
- *Influence:* This feature indicates the average impact of the news published by the user on social media. In other words, how many social media users receive the news published by this user on average? This feature can easily be considered equal to the number of followers, although the influence of each follower can also be important in determining the user's influence.
- *Sociality:* This feature shows how much the user interacts with other users. Simply, this feature can be considered equivalent to the number of friends.
- *Partisan bias:* This feature indicates the user's political orientation.
- *Activity credibility:* In the news field, this feature indicates how much of the news published by that user was fake or real. This feature can be calculated from the user's activity history on social media.
- *Activity level:* This feature indicates the amount of user activity (such as comments, shares and likes) on the received news.

**Propagation-related features:** These features determine how the news is propagating on social media. There are different patterns in spreading fake and real news on social media [19]. So, by extracting features related to propagation patterns, such as depth and level in the fake news cascade [4], we can estimate the possibility that the news is fake.

**Action-related features:** There are some other features related to the actions performed on a received news by the users. For example, the liking rate or the comments polarity of a news article can provide useful information about the authenticity of the news. To use these features effectively, it is necessary to consider the credibility of the user who created the action, because, for example, positive polarity in a comment can create different meaning depending the user's credibility.

Using these features, the issue of fake news detection can be considered as a classification problem. According to the availability of the content and user-related features at the publication stage, utilizing these features does not delay the detection, while propagation and action-related features require time to be created and the result is a delayed detection.

## 3- Related works

In this section, a brief review of research on fake news detection is provided. Fake news detection methods generally use news content and/or social context information. News content features can be extracted from text and images as well as news sources such as authors and websites that write or publish the news. News textual information can be used to extract features related to writing style at different language levels [20], i.e., lexicon-level [12, 13, 21, 22], syntax-level [13], semantic-level [12] and discourse-level [23]. These features can be explicitly obtained using methods like n-grams [12], Bag-Of-Words (BOWs) [13], Part-Of-Speeches (POSs) [13], Linguistic Inquiry and Word Count (LIWC) [12, 14], Rhetorical Structure Theory (RST) [15], etc.; or implicitly using deep neural networks with word embedding (for example word2vec [24]) to extract appropriate latent features that have shown good performance [16, 23, 25, 26]. One of the most important networks in the text classification area is Hierarchical Attention Network (HAN) [27]. In this network, which is based on Gated Recurrent Units (GRUs), two levels of attention are used at the word-level and the sentence-level. Signhania et al. [16] have provided a version of HAN, called 3HAN, specifically for detecting fake news, in which a layer of attention has been added at the Headline-Body level. Recently, convolutional neural networks (CNNs) have been utilized successfully in fake news detection[25, 28]. Visual features extracted from visual elements such as images and videos have also been used alongside textual features to detect fake news [21, 29, 30]. Zhou et al. [31] used the relationship (similarity) between the textual and visual information in news articles to predict the authenticity. Sitaula et al. [32] evaluated the credibility of the news using authors and content and Baly et al. [33] detected fake news by their source websites. Also a deep diffusive network model has been used to learn the representations of news articles, creators and subjects simultaneously [22].

Moreover, the use of social context information to detect fake news has recently become very attractive [34]. For example, Vosoughi et al. [19] have shown that fake news spreads faster, farther and more widely than true news. Utilizing user comments to detect fake news has recently been considered as well. For example, Cui et al. [35] applied user comments to identify important sentences in the news body. However, due to the fact that the use of user comments causes delays in detecting fake news, recent research has focused on the issue of early detection by, for example, adversarial learning [21] and user response generating [36] and unsupervised detection [37, 38]. Other social context information, like user profiles [39] and social connections [40], have also been used. Sentiment analysis has also been applied to detect fake news [41, 42] and rumors [43].

In this paper, we examine the publishers' role in detecting fake news on social media and suggest a high accurate multi-modal framework with early detection capability.

## 4- The proposed framework

In this section, we introduce our proposed method to detect fake news on social media before the propagation stage. The method uses content-related and user-related features simultaneously in order to improve the overall performance. Among the user-related features that we introduced in the previous section, two more important features, i.e., *"activity credibility"* and *"influence"* of publishers, are considered in the proposed framework, however it is easy to add and use other features. The framework of the proposed method, namely FR-Detect (**F**ake-**R**eal **Detect**or), is illustrated in Figure 6. As shown, the framework consists of five main parts including *Latent Linguistic Features Extractor*, *Credit Assessor*, *Influence Assessor*, *Integrator* and *Classifier,* which are described in the following subsections.

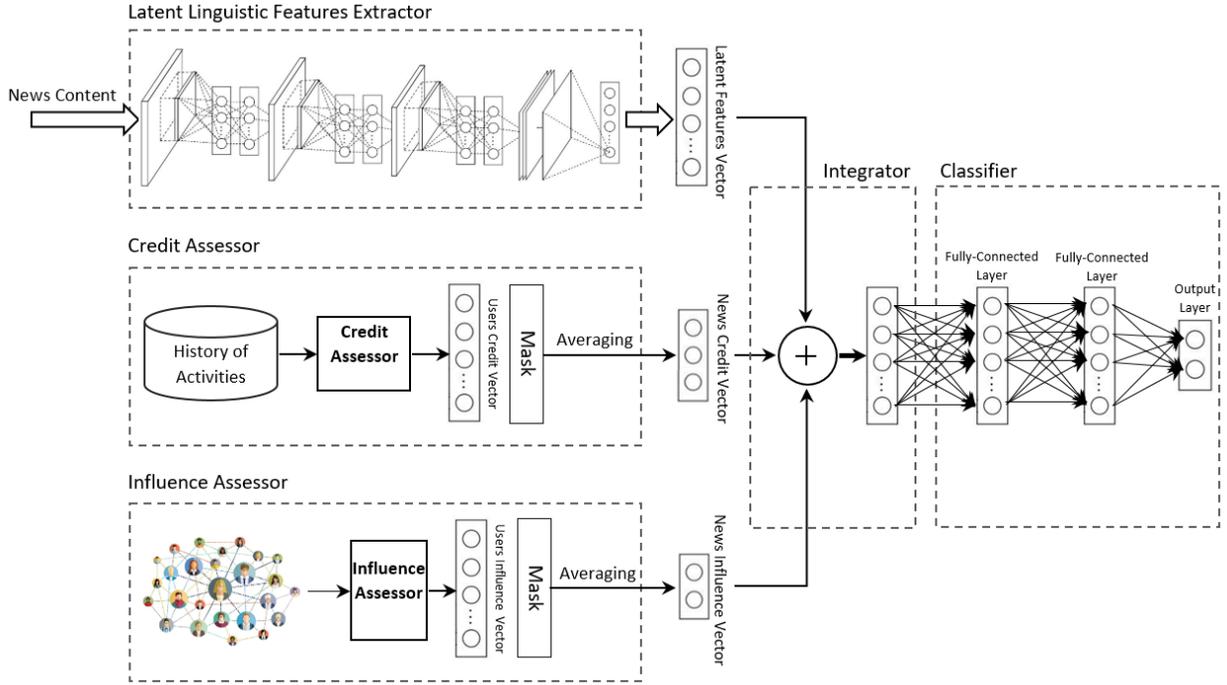

*Figure 6- The framework of FR-Detect.*

### 4-1- Latent Linguistic Features Extractor

Due to the importance of real-time representation and learning of content-related features in the scope of fake news detection, this part is designed based on deep learning methods. For this purpose, we have designed a sentence-level convolutional neural network (SLCNN). In this network, the news headline and body are transformed into a three-dimensional tensor, illustrated in Figure 7. As shown in the figure, the headline and the sentences of the body form the first dimension of the tensor. In the same way, the words of the sentences shape the second dimension, while the third dimension represents the word vectors of the words. The pre-trained word embedding, e.g., *word2vec* [24] or *GloVe* [44], could be used for representing the word vectors.

Since the input size of the network must be fixed, two thresholds are considered to adjust the different sizes of both texts and sentences (one for the number of sentences in the texts, $T_d$, and the other for the number of words in the sentences, $T_s$). The texts and the sentences longer than the thresholds would be cropped and shorter ones would be padded by zeros.

After some statistical analysis on the datasets in our experiments, as well as considering the structure of the SLCNN, we chose $T_s$=46 (about 2% of sentences have more than 46 words). In the same way, the threshold for the number of sentences in the news body is calculated by the following equation:

$$T_d = \lceil \mu + \sigma \rceil \tag{1}$$

where μ is the average number of sentences in the news body, and σ is the standard deviation. As a result, the performance of the model is significantly improved by ignoring the outlier sizes and preventing the construction of very large and sparse tensors.

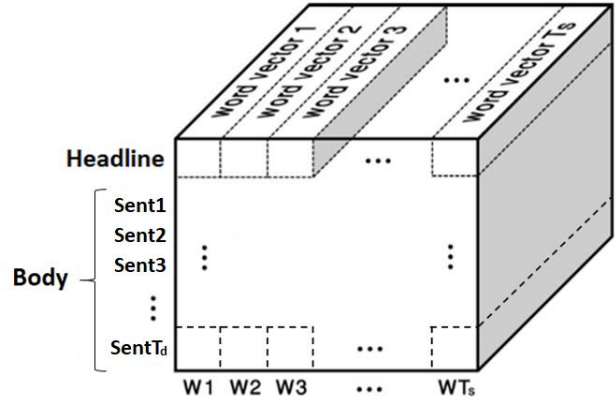

Figure 7- Shape of the transformed news content.

The architecture of the SLCNN is illustrated in Figure 8. Overall, in the input layer, the news articles are provided in the shape of the introduced 3D tensor. Then, using four horizontal convolutional blocks (HCB), one feature vector is extracted for each sentence individually. The main advantages of the SLCNN over traditional CNN for text classification [45] are: 1) the positional information of the sentences is used in the learning process, and 2) the SLCNN enables us to combine other extra features at the sentence level.

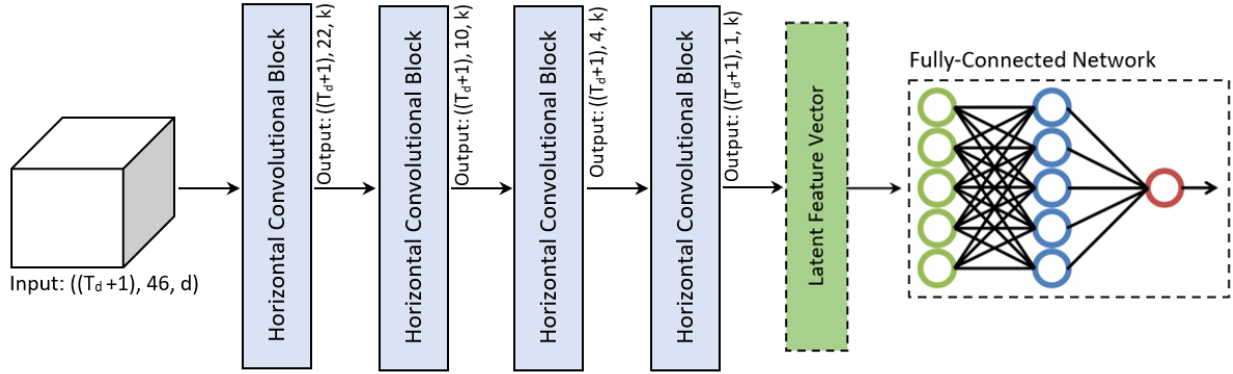

Figure 8- The architecture of the SLCNN.

Looking at the details of the HCB, as shown in Figure 9, there are two sequential convolution layers, each one followed by a Rectified Linear Unit (ReLU) activation function, $f(x)= max (0, x)$. A convolution operation consists of a filter $w \in \mathbb{R}^{s \times t \times d}$, which is applied to each possible window of $s \times t$ features from its input feature map, $X$, to produce a new feature map by equation 3:

$$X = \begin{bmatrix} x_{1,1} & x_{1,2} & \cdots & x_{1,n} \\ x_{2,1} & x_{2,2} & \cdots & x_{2,n} \\ \vdots & \vdots & & \vdots \\ x_{m,1} & x_{m,2} & \cdots & x_{m,n} \end{bmatrix} \quad (2)$$

$$\tilde{x}_{i,j} = f(w \cdot x_{i,j:i+s-1,j+t-1} + b) \tag{3}$$

where $x_{i,j:y,z}$ is the concatenation of features within the specified interval, $b \in \mathbb{R}$ is a bias term and $f$ is a non-linear function such as the ReLU. For our purpose, we consider s=1 and t=2. In the first convolution layer of the first HCB, d (the third dimension of the filters) is equal to the size of the word vectors, and in other cases d=1. At the end of the blocks, there is a max-pooling operation, with the pooling size = 2, that is applied over the generated intermediate feature map to select the maximum value from any two adjacent features as a more important feature. The new feature map is calculated by following equations:

$$\tilde{x}_{i,j} = \max\{x_{i,2j-1}, x_{i,2j}\} \tag{4}$$

The process of extracting one feature from one filter was described. The model uses multiple filters to obtain multiple features. The final extracted features are passed to the fully-connected layers (the Classifier) that end to a *softmax* output layer which is the probability distribution over labels.

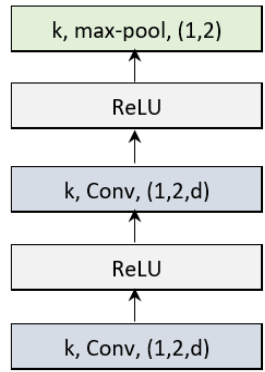

*Figure 9- The horizontal convolutional blocks. k is the number of filters.*

### 4-2- Credit Assessor

Due to the importance of publishers' credibility in determining the authenticity of the news, this module is responsible for calculating the news credit vector based on its publishers' credibility. As shown in Figure 6, this module determines the credibility activity of users based on their history in publishing news by pairs *(UCT, UCF)*, which *UCT* is the total number of news published by the user and *UCF* is the number of fake news published by the user. Then, the mask function selects the relevant publishers for the news article and creates the news credit vector *(NCT, NCF, numP)* by averaging, where *NCT* is the average number of news published by the news publishers, *NCF* is the average number of fake news published by the news publishers and *numP* is the number of the news publishers. All the values are normalized by min-max normalization.

### 4-3- Influence Assessor

As mentioned before, another important feature of the news publishers on social media is their reputation or influence. It means the news published by a more famous publisher can affect more users on social media. This feature also seems to be helpful in detecting fake news. By providing a definition and calculation formula for the publishers' influence on social media, its usefulness in detecting fake news has been investigated in the proposed FR-Detect framework.

*Definition (User influence on social media)*: user influence is the average impact of the news published by the user on social media.

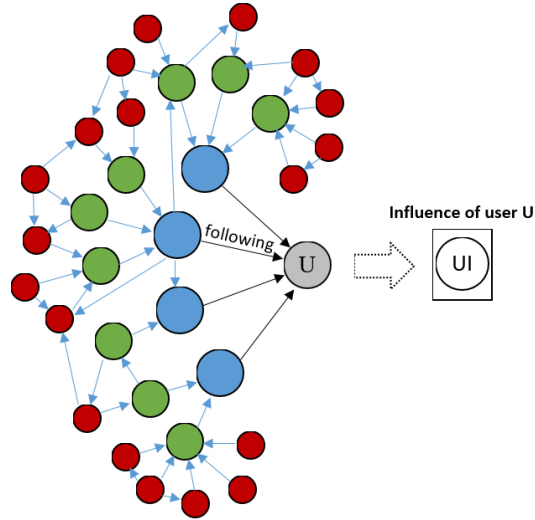

*Figure 10- Followers Network of user U. Blue users are Level 1 followers, Green users are Level 2 followers, and Red users are Level 3 followers of user U.*

According to the definition, the user's influence on social media is equal to the average ratio of users receiving the news published by that user on social media. Considering an example of a following network, shown in Figure 10, we propose the following equation to calculate a user's influence on social media:

$$\text{UI}(u) = \frac{1}{N-1}\left(|f_1(u)| + \sum_{i=2}^{d} p^{i-1}\left|f_i(u) - \bigcup_{j=1}^{i-1} f_j(u)\right|\right) \tag{5}$$

Where *N* is the total number of users on social media, *d* is the diameter of the network, *p* is the average probability of sharing news by users, and $f_i(u)$ is the set of the level *i* followers of user u on the network, which is calculated by the following equation:

$$f_i(u) = \begin{cases} \text{set of followers of u}, & i = 1 \\ \bigcup_{x \in f_{i-1}(u)} f_1(x), & i \geq 2 \end{cases} \tag{6}$$

For simplicity, the influence of users can be estimated by the number of followers. As shown in Figure 6, after calculating the user's influence, the mask function selects the relevant publishers for the news article and creates the news influence vector *(NI, numP)* by averaging, where *NI* is the average influence of the news publishers and *numP* is the number of the news publishers. All the values are normalized by min-max normalization.

**4-4- Integrator**

Once the desired features get ready, they need to be integrated to enter the classifier. As shown in Figure 11, the integrator concatenates explicit features (here the news credit and influence vectors) to the latent linguistic features at the sentence level. Then, one new feature vector with size *k* (k is equal to the number of filters) is extracted for each row using the proper number of the HCBs. Finally, the final integrated feature vector is prepared by flattening the vectors and sent to the classifier.

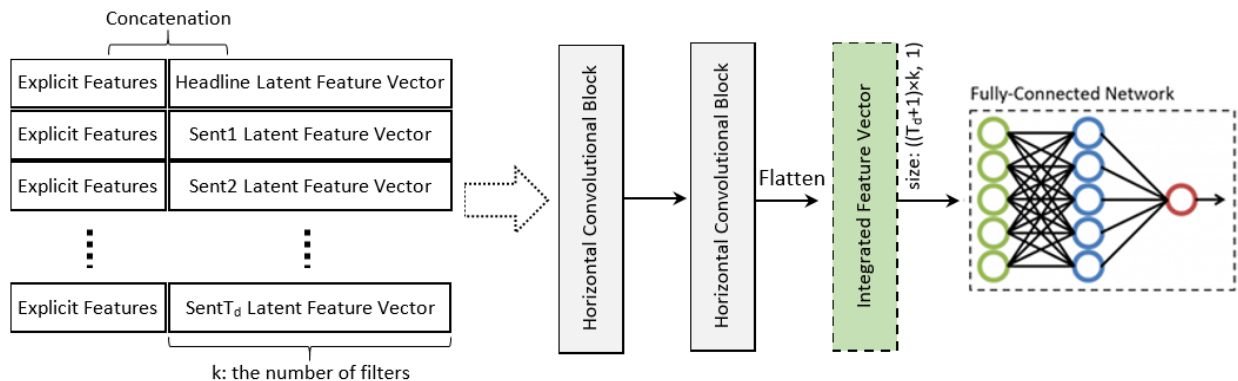

*Figure 11- The architecture of the Integrator.*

**4-5- Classifier**

Once the integrator integrates the required features, the Classifier is ready for learning and classifying the news articles based on provided features. This module includes two hidden fully-connected layers that end to a *softmax* output layer for classification. For regularization, a dropout module [46] is employed after each fully-connected layer.

**5- Experiments**

**5-1- Experimental settings**

In this section, we introduce the settings used in our experiments. For the SLCNN, the Natural Language Toolkit (NLTK) was used in order to tokenize words and sentences. In the input layer, as mentioned before, a pre-trained word-embedding is used to convert the words into the corresponding word vectors. The 100-dimensional GloVe vectors have been used in our experiments. Out-Of-Vocabulary (OOV) words are initialized from a uniform distribution with range [-0.01, 0.01]. We set the number of filters to 8 for all the convolutional blocks.

Due to the limitations of the available datasets, we considered the number of followers as the influence of the publishers. All the values for the news credit vector and the news influence vector are normalized using the min-max normalization method. We have also set the size of the fully-connected layers to 64, and both the dropout rates are set to 0.5. The model's parameters were trained by the Adam Optimizer [47], with the initial learning rate of 0.001.

**5-2- Benchmark datasets**

Due to the need for social context data along with news content to conduct our experiments, we utilize one of the comprehensive fake news detection benchmark dataset called *FakeNewsNet* [48]. The dataset is collected from two fact-checking platforms: *GossipCop* (news related to celebrities) and *PolitiFact* (political news), both containing labeled news content and related social context information in Twitter. The detailed statistics of the datasets are shown in Table 1.

*Table 1- Statistics of the datasets.*

| Platform | PolitiFact | | GossipCop | |
|---|---|---|---|---|
| | Real | Fake | Real | Fake |
| # Train samples | 192 | 188 | 9342 | 3162 |
| # Test samples | 49 | 47 | 2336 | 790 |
| $T_d$ | 280 | | 85 | |
| # Publishers | 512370 | | | |

**5-3- Results**

To evaluate the performance of fake news detection methods, we use the following metrics, which are commonly used to evaluate classifiers in related areas: Accuracy, Precision, Recall, and F1. First, we compare the performance of the SLCNN (as our base model) with traditional CNN for text classification [45]. As shown in Figure 12, the SLCNN has been able to achieve significantly better results than the text-CNN in all metrics due to having more information from the text.

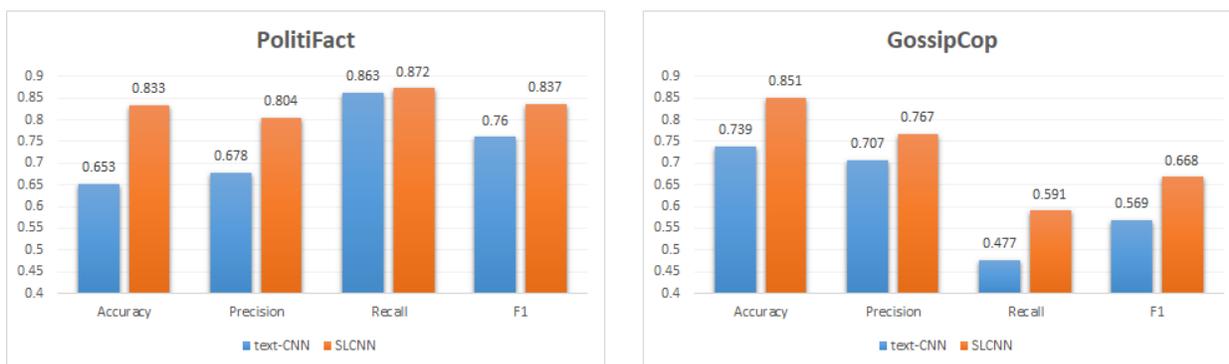

*Figure 12- Comparison of the performance of the SLCNN and the text-CNN.*

In order to evaluate the effectiveness of each module (publishers' feature), i.e., *Credit Assessor* and *Influence Assessor*, on performance of fake news detection models, we have developed four variants of FR-Detect framework:

- *FR-Detect (SLCNN):* FR-Detect (SLCNN) is the base version of FR-Detect without considering any user-related features. In other words, FR-Detect (SLCNN) uses only the news content.
- *FR-Detect (SLCNN, C):* FR-Detect (SLCNN, C) is a variant of FR-Detect that uses the *Credit Assessor* module as well as the news content. In other words, we remove the *Influence Assessor* from FR-Detect.

- *FR-Detect (SLCNN, I):* FR-Detect (SLCNN, I) is a version of FR-Detect that uses the *Influence Assessor* module as well as the news content. In other words, we remove the *Credit Assessor* from FR-Detect.
- *FR-Detect:* FR-Detect is the full version of FR-Detect framework, which is introduced in Section 4.

*Table 2- Module analysis for FR-Detect framework.*

| Datasets | Metrics | FR-Detect (SLCNN) | FR-Detect (SLCNN,C) | FR-Detect (SLCNN,I) | FR-Detect |
|---|---|---|---|---|---|
| **PolitiFact** | Accuracy | 0.833 | 0.854 | 0.813 | **0.906** |
|  | Precision | 0.804 | 0.891 | 0.784 | **0.896** |
|  | Recall | 0.872 | 0.796 | 0.851 | **0.915** |
|  | F1 | 0.837 | 0.841 | 0.816 | **0.905** |
| **GossipCop** | Accuracy | 0.851 | 0.974 | 0.883 | **0.978** |
|  | Precision | 0.767 | 0.933 | 0.774 | **0.964** |
|  | Recall | 0.591 | **0.968** | 0.761 | 0.948 |
|  | F1 | 0.668 | 0.950 | 0.767 | **0.956** |

The performance analysis for the variants of FR-Detect are summarized in Table 3 and compared in Figure 13. From the results, we make the following observations:

- When we add the *Activity Credibility* information of news publishers to the SLCNN, the results improve dramatically across almost all metrics in both datasets. This indicates that the credibility of publishers plays a crucial role in verifying the authenticity of the news.
- When we add the *Influence* information of news publishers to the SLCNN, we see remarkable performance improvement in *GossipCop*, while the results are slightly reduced in *Politifact*.
- FR-Detect has achieved the best results in both datasets. In *GossipCop*, this improvement is small compared to FR-Detect (SLCNN, C), while in *PolitiFact* it is very significant, i.e., about 7% in accuracy and F1. In case of *PolitiFact*, this amount of improvement can be attributed to the high Precision in FR-Detect (SLCNN, C) and the high Recall in FR-Detect (SLCNN, I). In other words, FR-Detect (SLCNN, C) was able to detect fake news with less error, whereas FR-Detect (SLCNN, I) was able to detect a greater proportion of fake news.

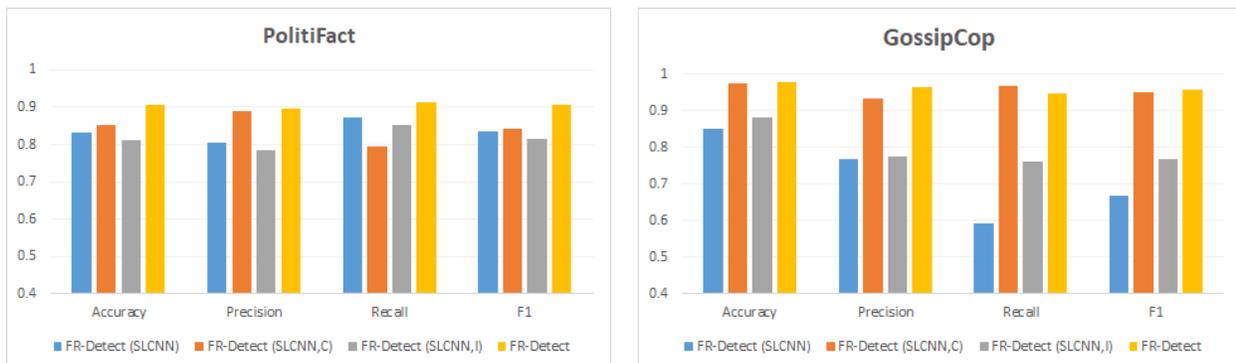

*Figure 13- - The performance comparison of modules in FR-Detect.*

We also compared the performance of FR-Detect with state-of-the-art methods for fake news detection. The algorithms used for comparison are listed as follows:

- **3HAN** [16]: 3HAN utilizes a hierarchical attention neural network framework on news textual contents for fake news detection. It encodes textual contents using a three level hierarchical attention network, each for words, sentences, and the headline.
- **TCNN-URG** [36]: TCNN-URG utilizes a Two Level Convolutional Neural Network with User Response Generator (TCNN-URG) where TCNN captures semantic information from textual content by representing it at the sentence and word level, and URG learns a generative model of user response to news contents from historical user responses in order to generate responses for new incoming articles and use them in fake news detection.
- **dEFEND** [35]: dEFEND utilizes a sentence-comment co-attention sub-network to exploit both news contents and user comments to jointly capture top-k check-worthy sentences and user comments for fake news detection.
- **SAFE** [31]: SAFE uses multi-modal (textual and visual) information of news articles. First, neural networks are adopted to separately extract textual and visual features for news representation. Then the relationship between the extracted features is investigated across modalities. Finally, news textual and visual representations along with their relationship are jointly learned and used to predict fake news.

*Table 3-Performmace of methods in fake news detection.*

| Datasets | Metrics | 3HAN | TCNN-URG | dEFEND | SAFE | FR-Detect |
|---|---|---|---|---|---|---|
| **PolitiFact** | Accuracy | 0.844 | 0.712 | 0.904 | 0.874 | **0.906** |
| | Precision | 0.825 | 0.711 | **0.902** | 0.889 | 0.896 |
| | Recall | 0.899 | 0.941 | **0.956** | 0.903 | 0.915 |
| | F1 | 0.860 | 0.810 | **0.928** | 0.896 | 0.905 |
| **GossipCop** | Accuracy | 0.750 | 0.736 | 0.808 | 0.838 | **0.978** |
| | Precision | 0.659 | 0.715 | 0.729 | 0.857 | **0.964** |
| | Recall | 0.695 | 0.521 | 0.782 | 0.937 | **0.948** |
| | F1 | 0.677 | 0.603 | 0.755 | 0.895 | **0.956** |

Note that all the models used in this comparison, except dEFEND (because of using real comments), have the early detection property. The results are shown in Table 3 and compared in Figure 14. From the results, we make the following observations:

- Considering PolitiFact, although FR-Detect and dEFEND have shown better performance in Accuracy and F1-score respectively, but the difference has been relatively small for models dEFEND, SAFE and FR-Detect, which can be ignored due to the small number of samples in this dataset. We can sort the models as follows: dEFEND ≥ FR-Detect > SAFE > 3HAN > TCNN-URG.
- Regarding GossipCop, FR-Detect has managed to achieve by far the best result. Improvements in Accuracy and F1-score have been more than 14% and 6%, respectively. According to the results, we can sort the models as follows: FR-Detect > SAFE > dEFEND > 3HAN > TCNN-URG.

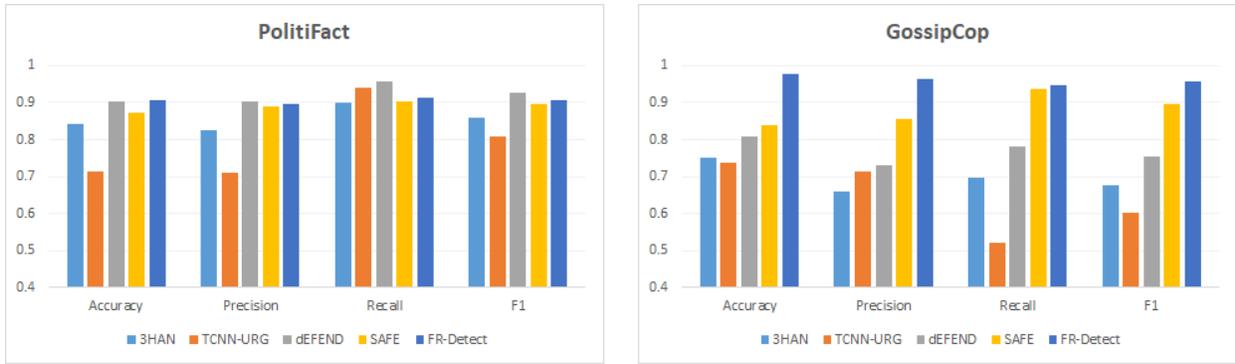

*Figure 14- Performance comparison of fake news detection methods.*

**5-4- Discussion**

In this section, we discuss two issues:

- **I1.** Cold start issue in the user-related features
- **I2.** Statistical information of the publishers' features according to whether the news is fake or real

Obviously, some user-related features have the cold start issue. This means that little information may be available about that feature because the user is a newcomer. This is also true for the two features of *Credibility* and *Influence*. Due to lack of significant number of followers of the newcomers, this issue is not important in terms of *Influence*, because the published news of these publishers cannot be widely disseminated on social media and therefore will not have much impact. But in terms of the credibility, this issue can affect the results. To evaluate the effectiveness of the cold start issue related to the credibility feature, we changed *NCT* and *NCF* to zero for 10%, 20% and 30% of the news in both the train and the test data randomly and uniformly. The results for *FR-Detect (SLCNN, C)* are illustrated in Figure 15 in terms of accuracy and F1. As shown, the cold start issue can reduce the accuracy of the results, but due to the simultaneous use of content-related features, this reduction is far less than the reduction of credibility-related information. In other words, combining the credibility feature with the news content features has significantly reduced the effect of the cold start problem. Note that new users make up a small proportion of social media users.

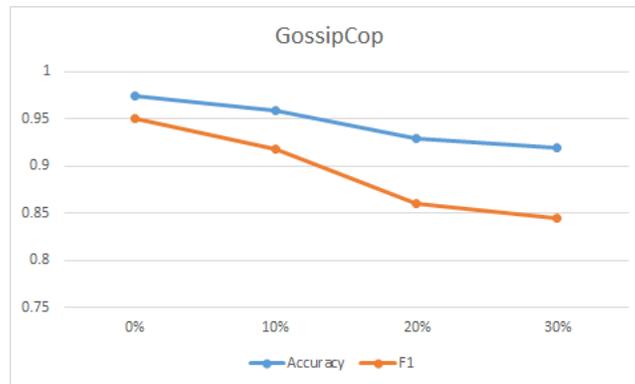

*Figure 15- The effect of the cold start issue in terms of publishers' credibility feature.*

The following results can be obtained by examining the statistical graphs related to the social context features used in FR-Detect, which are given for *PolitiFact* and *GossipCop* in Figures 16 and 17 respectively:

- Regarding the average number of published news (NCT), publishers of real news related to celebrities have published more news on average in the past, while in the political news, there is not much difference.
- Regarding the average number of published fake news (NCF), publishers of fake news in both datasets have published significantly more fake news in the past on average. By calculating NCF/NCT, this difference becomes more obvious. In other words, the news of publishers who have had more fake news in the past is more likely to be fake.
- Regarding the average number of followers (NI), fake news publishers had relatively fewer followers.
- Regarding the number of publishers (numP), it can be seen that real news has relatively more publishers than fake news.

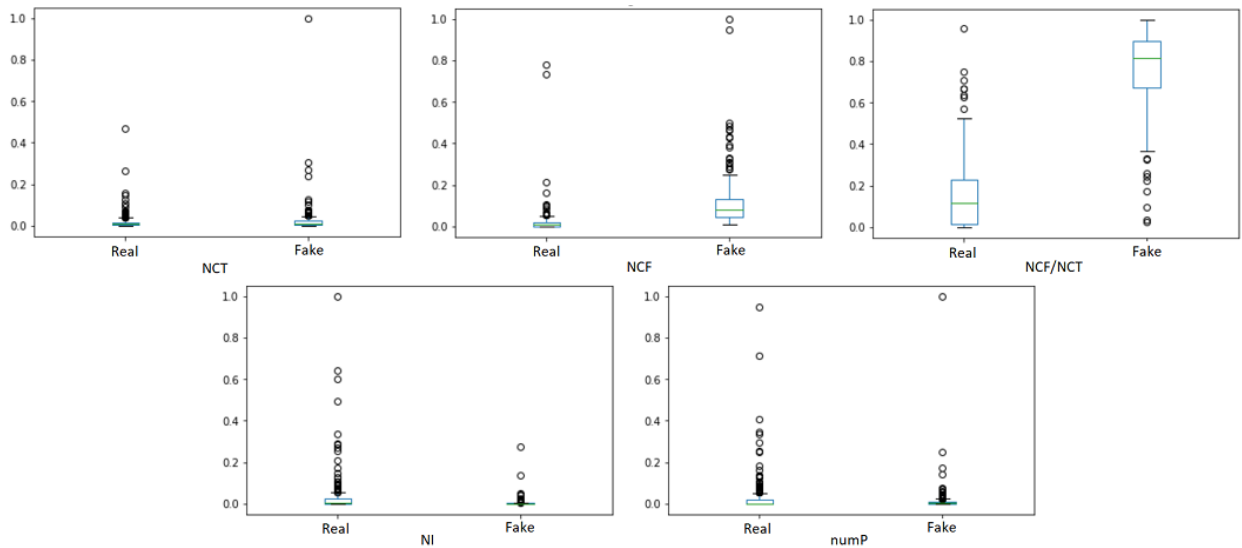

*Figure 16- The statistical graphs related to the social context features used in FR-Detect for PolitiFact.*

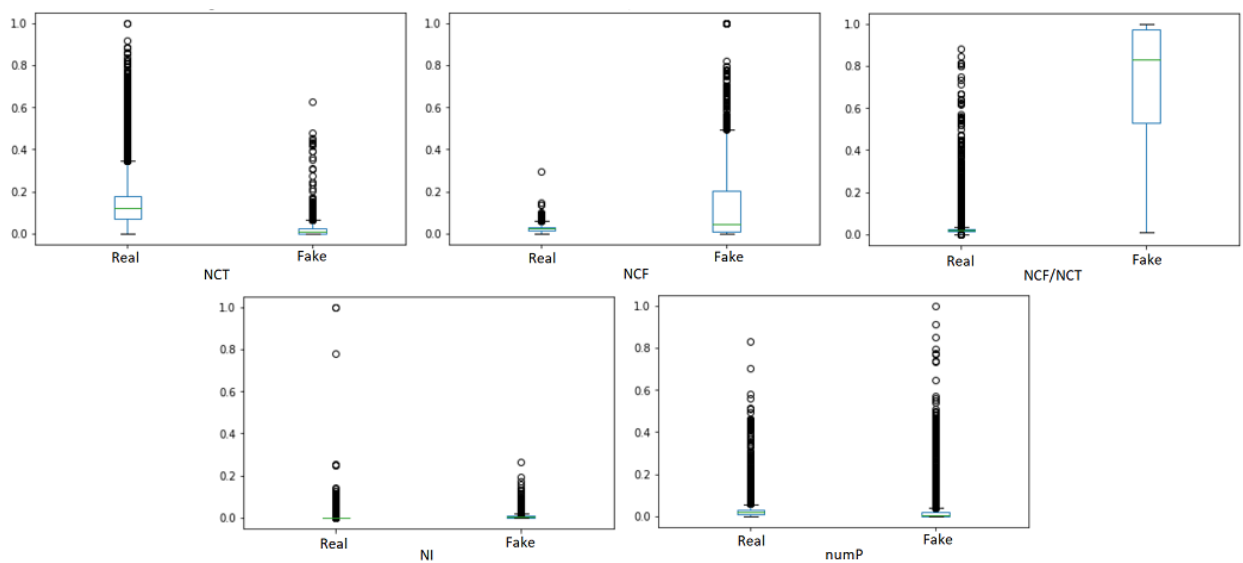

*Figure 17- The statistical graphs related to the social context features used in FR-Detect for GossipCop.*

## 6- Conclusion and future works

Fake news detection has received growing attention in recent years. In this paper, we introduced some important features including *Activity Credibility* and *Influence* for news publishers on social media and studied their role in detecting fake news. Experiments on real-world datasets demonstrate that the historical performance of publishers in publishing news can be well used in estimating the probability of the news being fake. Therefore, publishers' features along with content features can play an important role in detecting fake news and improving the performance of models. One of the most important advantages of publishers' features is that they do not delay the detection process because they are available at the publication time. We have also presented a novel sentence-level convolutional neural network (SLCNN) that can be used generally in the text classification. Experimental results have shown the proposed framework (FR-Detect) outperforms the state-of-the-art methods. In other words, FR-Detect has succeeded in detecting fake news with 97.8% and 95.6% in Accuracy and F1-score, respectively. As future work, we intend to extract and study more features from publishers and their interconnections.